\documentclass[10pt,letterpaper]{article}
\usepackage[top=0.85in,left=2.75in,footskip=0.75in]{geometry}

\usepackage{amsmath,amssymb}

\usepackage{changepage}

\usepackage{textcomp,marvosym}

\usepackage{cite}

\usepackage{nameref,hyperref}

\usepackage[nopatch=eqnum]{microtype}
\DisableLigatures[f]{encoding = *, family = * }

\usepackage[table]{xcolor}

\usepackage{array}

\newcolumntype{+}{!{\vrule width 2pt}}

\usepackage{algorithm}
\usepackage{algpseudocode}

\newlength\savedwidth

\raggedright
\usepackage[aboveskip=1pt,labelfont=bf,labelsep=period,justification=raggedright,singlelinecheck=off]{caption}

\makeatletter
\renewcommand{\@biblabel}[1]{\quad#1.}
\makeatother

\usepackage{lastpage,fancyhdr,graphicx}
\usepackage{epstopdf}
\fancyheadoffset[L]{2.25in}
\fancyfootoffset[L]{2.25in}
\begin{document}
\vspace*{0.2in}

\begin{flushleft}
{\Large
\textbf\newline{We Are What We Buy: Extracting urban lifestyles using large-scale delivery records} 
}
\newline
Minjin Lee\textsuperscript{1},
Hokyun Kim\textsuperscript{2},
Bogang Jun\textsuperscript{1,3,4},
Jaehyuk Park\textsuperscript{5*}
\\
\bigskip
\textbf{1} Research Center for Small Business Ecosystem, Inha University, Incheon, Republic of Korea
\\
\textbf{2} Department Brain and Cognitive Engineering, Korea University, Seoul, Republic of Korea
\\
\textbf{3} Department of Data Science, Inha University, Incheon, Republic of Korea
\\
\textbf{4} Department of Economics, Inha University, Incheon, Republic of Korea
\\
\textbf{5} KDI School of Public Policy and Management, Sejong, Republic of Korea
\\
\bigskip
* Corresponding Author: jp@kdischool.ac.kr
\\







\end{flushleft}
\section*{Abstract}
Lifestyle has been used as a lens to characterize a society and its people within, which includes their social status, consumption habits, values, and cultural interests.  
Recently, the increasing availability of large-scale purchasing records, such as credit card transaction data, has enabled data-driven studies to capture lifestyles through consumption behavior. 
However, the lack of detailed information on individual purchases prevents researchers from constructing a precise representation of lifestyle structures through the consumption pattern. 
Here, we extract urban lifestyle patterns as a composition of fine-grained product categories that are significantly consumed together. Leveraging 103,342,186 package delivery records from 2018 to 2022 in Seoul, Republic of Korea, we construct a co-consumption network of detailed product categories and systematically identify lifestyles as clusters in the network. 
Our results reveal five lifestyle clusters --- \textit{Beauty lovers}, \textit{Fashion lovers}, \textit{Work and life}, \textit{Homemakers}, and \textit{Baby and hobbyists} --- representing distinctive lifestyles, while also being connected to each other. Moreover, the geospatial distribution of lifestyle clusters aligns with regional characteristics (business vs. residential areas) and is associated with multiple demographic characteristics of residents, such as income, birth rate, and age. 
Temporal analysis further demonstrates that lifestyle patterns evolve in response to external disruptions, such as COVID-19. 
As urban societies become more multi-faceted, our framework provide a powerful tool for researchers, policymakers, and businesses to understand the shifting dynamics of contemporary lifestyles.


\section*{Introduction}

Lifestyle is a broad and multidimensional concept that reflects an individual's social status, consumption habits, values, and cultural interests\cite{zukin1998urban, bourdieu1984distinction}. It not only reflects the identity of individuals and groups but also serves as a lens through which we analyze social phenomena and understand the underlying mechanisms that shape them\cite{bellah1985habits, kozlowski2019}. Certain lifestyle patterns are intuitive and easy to identify. For example, a financial professional may exhibit a stereotypical lifestyle that includes frequent commuting, business travel, and active engagement with economic news. However, many lifestyle patterns are far less apparent. A growing body of research on ``lifestyle politics" has revealed unexpected correlations between cultural preferences (e.g., movies, music) and political orientations, which at first glance may appear unrelated~\cite{Posta2015latte, shi2017, Praet2021lifestylepoliticsmodeling}. These findings highlight the complex, multifaceted nature of lifestyle and the need to examine a wide range of elements to construct a more holistic understanding.

At the same time, lifestyles are continuously reshaped and evolved through complex interactions between individuals and their social environments~\cite{bellah1985habits, boy2020instagram, low2003gated}. Individuals observe others’ consumption behaviors, learn from them, and express their interests and identities through their own consumption patterns. Hence, understanding what people consume has become a crucial means of analyzing lifestyle structures in modern society \cite{florida2002rise, zukin1995cultures, low2003gated}. Recognizing the strong link between lifestyle and consumption, previous research has sought to analyze lifestyle through consumption data using various approaches. 

More recently, data-driven studies have attempted to analyze lifestyle through digital consumption footprints, leveraging credit card transactions and mobility data~\cite{di2018sequences}. Although these studies open up a computational data-driven approach in lifestyle research, they have to focus on \textit{where} people consume (location-based data) in a broader sense, rather than \textit{what} they consume (product-based data), due to the lack of detailed information on the consumed product. This limits the ability to capture the detailed composition of lifestyles, since broader consumption categories obscure the granular distinctions that define different lifestyle groups.

To challenge these limitations, this study proposes a new approach to lifestyle analysis.
We leverage a highly detailed dataset consisting of 103,342,186 package delivery records from Seoul, Republic of Korea, spanning five years from 2018 to 2022. This dataset enables an unprecedented level of granularity in capturing consumer behavior, allowing us to systematically detect and classify lifestyle patterns based on real-world consumption habits. Our result find five lifestyle patterns at the top-level of our lifestyle hierarchy, which provides a broad overview of lifestyle categories and detailed insights into specific trends. 
We also examine how lifestyle patterns have evolved, particularly in response to COVID-19, and discuss their broader social implications.  

\section*{Results}

\begin{figure}[!ht]
\centering
\includegraphics[width=13cm]{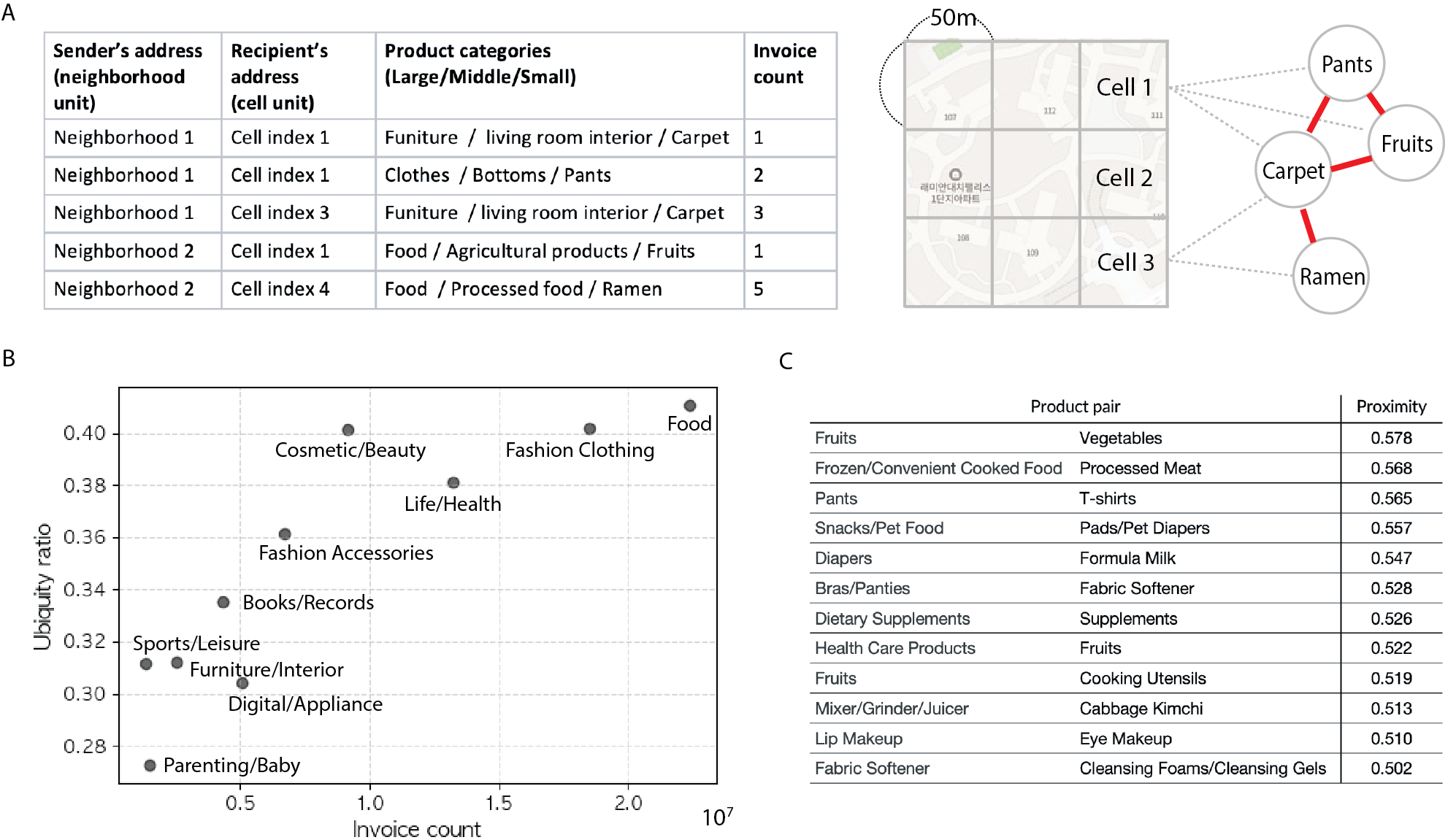}
\caption{{\bf Data Description and Consumption Patterns}  
A: Structure of the dataset and an overview of the co-consumption network construction.  
B: Invoice count and normalized ubiquity for major product categories.  
C: Examples of product pairs with a high proximity.}
\label{fig1}
\end{figure}

Here, we focus on product sets that are frequently delivered together, constructing a co-consumption network to identify commonly co-purchased product baskets. These product baskets represent grouped consumer interests, forming the core element sets that define lifestyles. 

We use package delivery records at high spatial resolution (50m × 50m grid cells) for five years (2018-2022) in Seoul. The dataset consists of five key attributes: sender location, receiver location, product categories, and the number of delivery invoices, as shown in Fig. \ref{fig1}A. Product categories are classified into three levels of categories hierarchically, where 533 product categories exist at the finest level. 
To determine overrepresented products in a given cell, we compute the Revealed Comparative Advantage (RCA)~\cite{hidalgo2007product}. The RCA for product \( p \) in grid cellw\( c \) is defined as:

\begin{equation}
\text{RCA}_{p,c} = 
\left( \frac{X_{p,c}}{\sum_p X_{p,c}} \right) \Bigg/ \left( \frac{\sum_c X_{p,c}}{\sum_{p,c} X_{p,c}} \right)
\end{equation}

where \( X_{p,c} \) represents the purchase volume of product \( p \) in grid cell \( c \). Only products with RCA values greater than 1 are considered meaningful purchases and are included in the network construction. Next, we measure the connection strength between products by calculating their proximity, which represents the conditional probability that both products are co-purchased within the same grid cell. The proximity between products \( p \) and \( p' \) is given by:

\begin{equation}
\phi_{p,p'} = \min \left( P(p | p'), P(p' | p) \right)
\end{equation}

where 

\begin{equation}
\quad P(p | p') = \frac{X_{p,p'}}{X_{p'}}, \quad
P(p' | p) = \frac{X_{p,p'}}{X_{p}}
\end{equation}
Here, \( X_{p,p'} \) represents the number of times products \( p \) and \( p' \) are co-purchased in the same grid cell, while \( X_{p} \) and \( X_{p'} \) denote the total purchases of products \( p \) and \( p' \), respectively.

\subsection*{Urban Consumption Patterns}

We first examine the general consumption patterns observed in the dataset. Fig. \ref{fig1}B presents two key aspects at the large-category level: (1) the total number of delivered products (invoice count) and (2) their distribution across various locations (ubiquity). Ubiquity measures the number of grid cells where a given product has an RCA greater than 1, indicating how widely it is consumed. Highly ubiquitous products are broadly distributed, whereas less ubiquitous products tend to be specialized and localized. For better interpretability, we normalize ubiquity by dividing the ubiquity count by the total number of grid cells, representing the proportion of areas where each product is significantly purchased. Overall, invoice count and ubiquity exhibit a positive correlation. Essential goods such as food and clothing are the most frequently and widely consumed, followed by life/health products and cosmetics/beauty items. In contrast, categories such as sports/leisure, parenting/baby, and furniture/interior have significantly lower invoice counts and ubiquity ratios, suggesting they are consumed by more specific demographic groups or purchased less frequently.

Fig. \ref{fig1}C provides examples of the proximity of product pairs. The displayed product pairs have high proximity values (greater than 0.5), whereas most product proximities range between 0.1 and 0.45. The identified highly correlated products typically exhibit complementary relationships~\cite{neffke2019value}. Some pairs, such as fruits and vegetables, pants and T-shirts, diapers and formula milk, and lip makeup and eye makeup, intuitively reflect clear complementarities. However, other pairs, such as mixer/grinder/juicer and kimchi, fruits and cooking utensils, and fabric softener and cleaning products, may not immediately appear complementary. Although detecting pairs of complementary products is not the main objective of this study, examining their relationships with other products allows us to uncover complementary contexts that might not be evident in direct pairwise comparisons.

\subsection*{Lifestyle Patterns from Co-consumption Network}

\begin{figure}[!h]
\centering
\includegraphics[width=13cm]{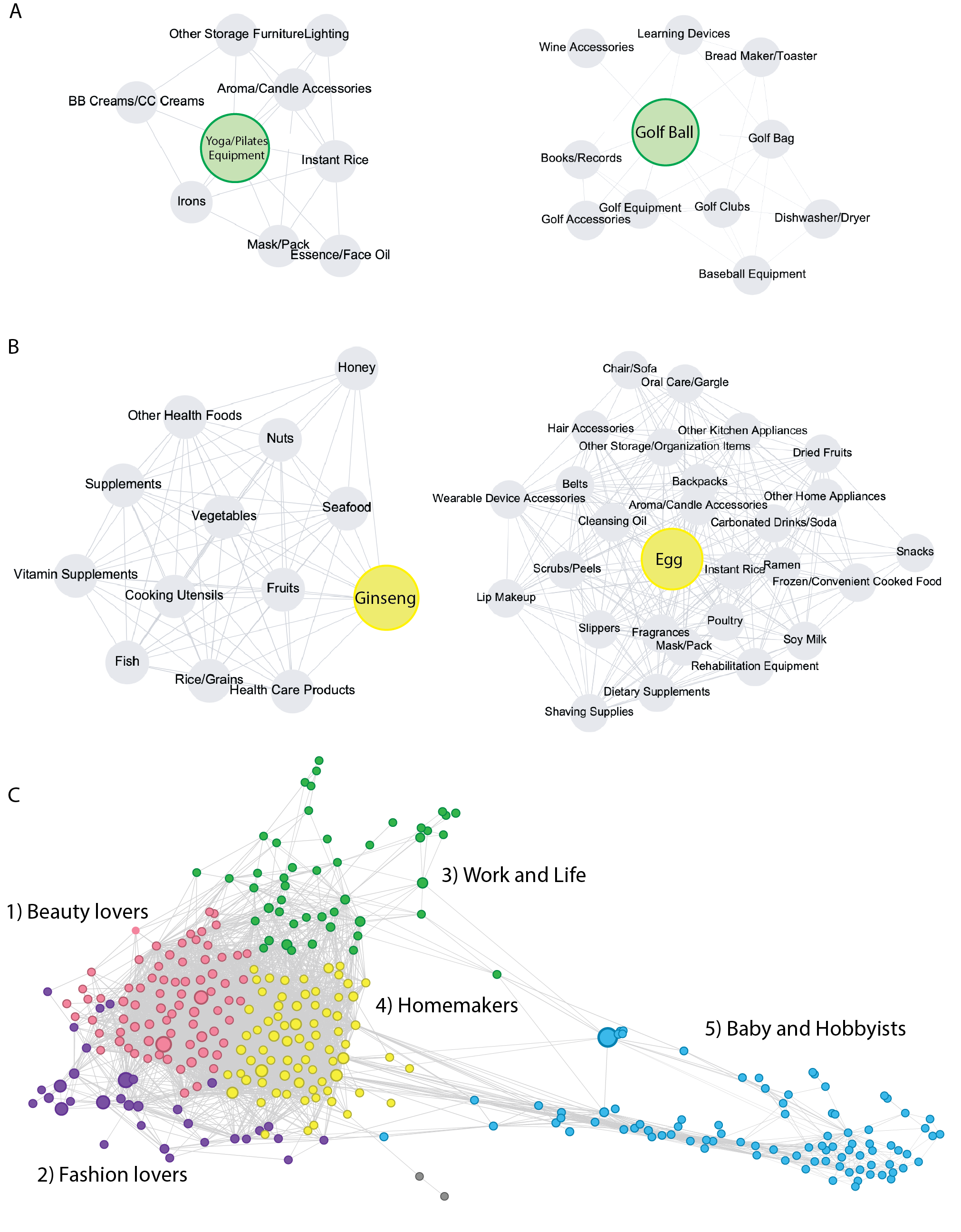}
\caption{{\bf Co-consumption network}
A: Ego network examples of sports items (Yoga/Pilates equipment and Golf ball). B: Ego network examples of food items (Ginseng and Egg). C: Co-consumption network with community division. The colors of the nodes represent the communities they belong to: 1) \textit{Beauty lovers} (Pink), 2) \textit{Fashion lovers} (Purple), 3) \textit{Work and life} (Green), 4) \textit{Homemakers} (Yellow), and 5) \textit{Baby and hobbyists} (Blue).}
\label{fig2}
\end{figure}

Our co-consumption network by integrating pairwise proximity information reveals complex consumption patterns and latent lifestyles (For a detailed information about the network generation and filtering process, see Materials and method).
We examine how the network representation reflects consumption contexts and consumer perceptions. Fig. \ref{fig2}A and B illustrate examples of product consumption contexts. Fig. \ref{fig2}A presents ego networks of two sports-related items: Yoga and Pilates equipment and Golf balls. The yoga and pilates equipment network (left) exhibits strong co-consumption with cosmetic products, home styling interior decorations, and instant rice, indicating that yoga and pilates are perceived as part of a broader self-care– and aesthetics-oriented lifestyle, rather than merely as sports activities. In contrast, the golf ball network (right) highlights strong connections with other hobby-related products. Consumers who purchase golf balls also tend to buy golf-related equipment, baseball gear, books, and learning devices. The presence of books and learning devices in this network suggests that golf enthusiasts may have a strong interest in self-improvement, whether through refining their sports skills or expanding their knowledge in other areas.

Fig. \ref{fig2}B explores the consumption contexts of ginseng and eggs. The ginseng ego network reveals multiple connections with health foods, food ingredients, and cooking tools, reinforcing the common perception of ginseng as a health-promoting food. On the other hand, the egg network is linked to a broader and more diverse range of products. Within the food category, eggs are closely associated with instant rice, carbonated drinks, ramen, and frozen meals, indicating a strong connection to convenience foods and diet-related consumption. In addition, eggs are frequently purchased alongside beauty products and interior items, such as aroma candles, cleansing oils, and fragrances, suggesting that consumers who prioritize self-care and beauty may also prefer eggs as part of their lifestyle choices. While eggs are traditionally considered a fundamental cooking ingredient, their consumption patterns in the network indicate that they are increasingly perceived as a convenience food rather than a core element of home-cooked meals. These findings illustrate how a co-consumption network can reveal consumer perceptions of products and reflect broader lifestyle trends.

In order to systematically uncover the underlying structure of urban lifestyles, we extract the community structure of the network using the Louvain algorithm, which detects communities by maximizing modularity --- a measure of the strength of division in a network~\cite{newman2006modularity, blondel2008fast}. To mitigate algorithm-induced uncertainty, we iteratively run the algorithm to determine the most stable community division\cite{Cho2023,lee2021inconsistence} (For a detailed description, see Materials and methods). As shown in Fig. \ref{fig2}C, the optimal community structure segments the network into five well-defined lifestyle clusters. Based on the product compositions within each community, we label them as representing the following lifestyle clusters: \textit{(1) Beauty lovers}, \textit{(2) Fashion lovers}, \textit{(3) Work and life}, \textit{(4) Homemakers}, and \textit{(5) Baby and hobbyists}.

Each community is characterized as follows: \textit{(1) Beauty lovers} includes various cosmetic products, hair care and bath items, fashion accessories such as bags and shoes, and fitness-related products like massage tools and fitness accessories. \textit{(2) Fashion lovers} encompass all types of outerwear and clothing, diverse footwear, fashion accessories such as rings, and hair styling products like dye solutions. \textit{(3) Work and life} consists of PC-related products, stationery and office supplies, convenience foods, snacks, coffee/tea, and some office furniture such as tables and chairs. \textit{(4) Homemakers} include various food ingredients, cleaning and laundry supplies, cooking utensils, basic clothing, toilet paper, and functional apparel and cosmetics related to anti-aging, such as shapewear, neck care products, and sunscreen. These four communities are distinct but densely interconnected within a larger network. On the other hand, \textit{(5) Baby and hobbyists} includes a broad range of baby and parenting products alongside various hobby and sports-related items. This community is clearly separated from the other four, indicating that parenting and hobby-related lifestyles represent distinct domains compared to more generalized lifestyle categories.

\begin{figure}[!h]
\centering
\includegraphics[width=13cm]{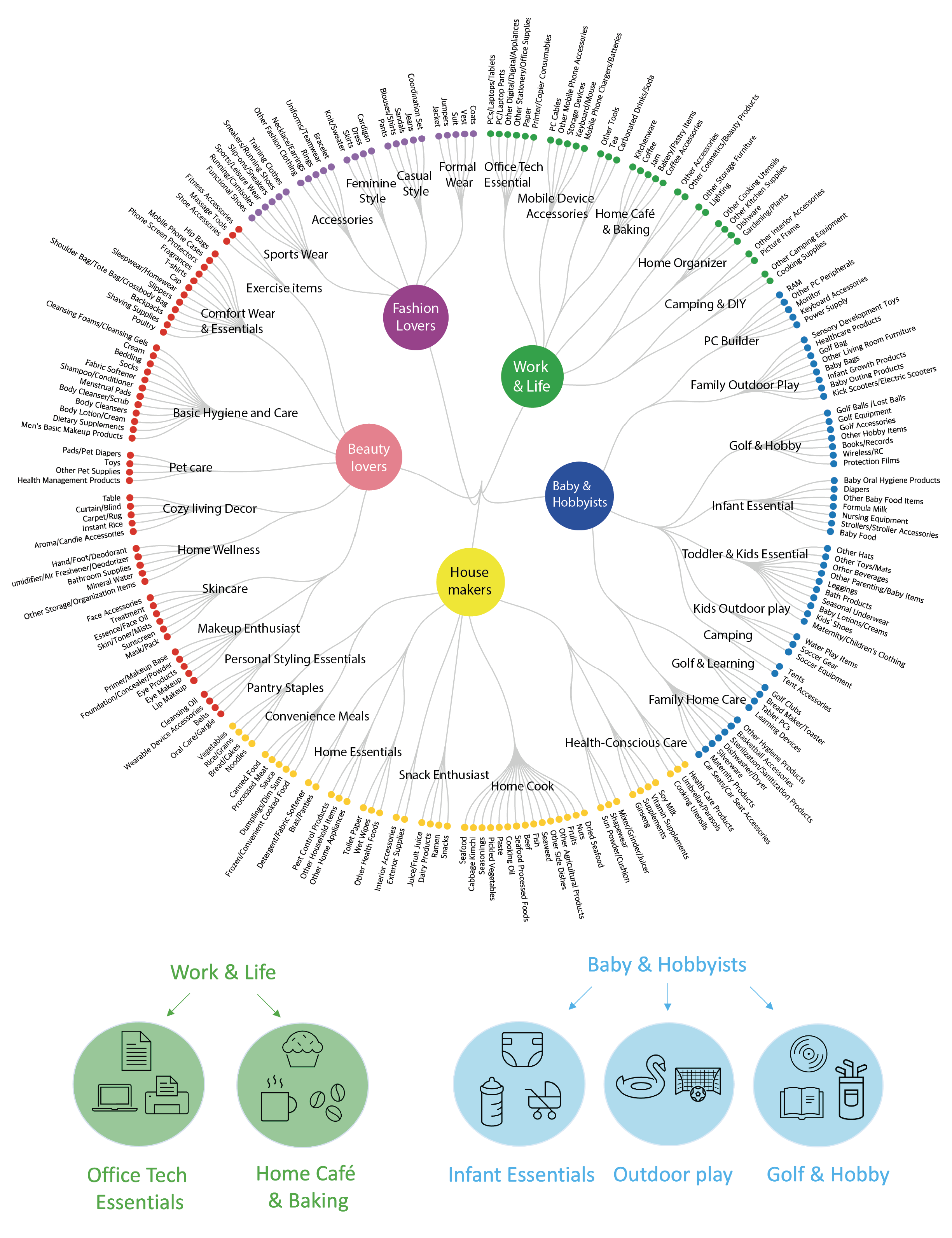}
\caption{{\bf Hierarchical structure of urban lifestyle}
The large circles in the middle represent
the high level of lifestyle clusters and tree branches show the lower level of detailed lifestyles. We also name each lifestyle cluster based on the products in the cluster.}
\label{fig3}
\end{figure}

To explore the community structure at finer resolutions, we further apply hierarchical community detection, iterating the algorithm within each identified community\cite{park2019global}. The subdivision process continues until the modularity score remains above 0.25, ultimately revealing a three-level hierarchical structure (Fig. \ref{fig3}). While first-level communities provide a broad yet somewhat ambiguous classification due to the inclusion of diverse product categories, second- and third-level communities offer a much clearer segmentation of lifestyles. For example, the \textit{Baby and hobbyists} community is sharply divided into baby and parenting groups and hobby groups, which further split into specialized subgroups such as baby care products and outdoor play equipment. Additionally, distinct hobbyist communities emerge, such as the golf-book cluster, where golf-related equipment and books are frequently co-purchased, and the camping community, which includes tents and other outdoor gear. Health-conscious food communities, including ginseng and dietary supplements, are clearly distinguished from groups focused on convenience and processed foods. The \textit{Work and life} community also exhibits further specialization, forming distinct groups for office productivity, coffee breaks, and home decoration. These findings demonstrate that lifestyles are not arbitrary collections of preferences but rather structured, interconnected systems embedded within urban consumer behavior.

\subsection*{Socioeconomic Relationship in Lifestyle}

\begin{figure}[!h]
\centering
\includegraphics[width=13cm]{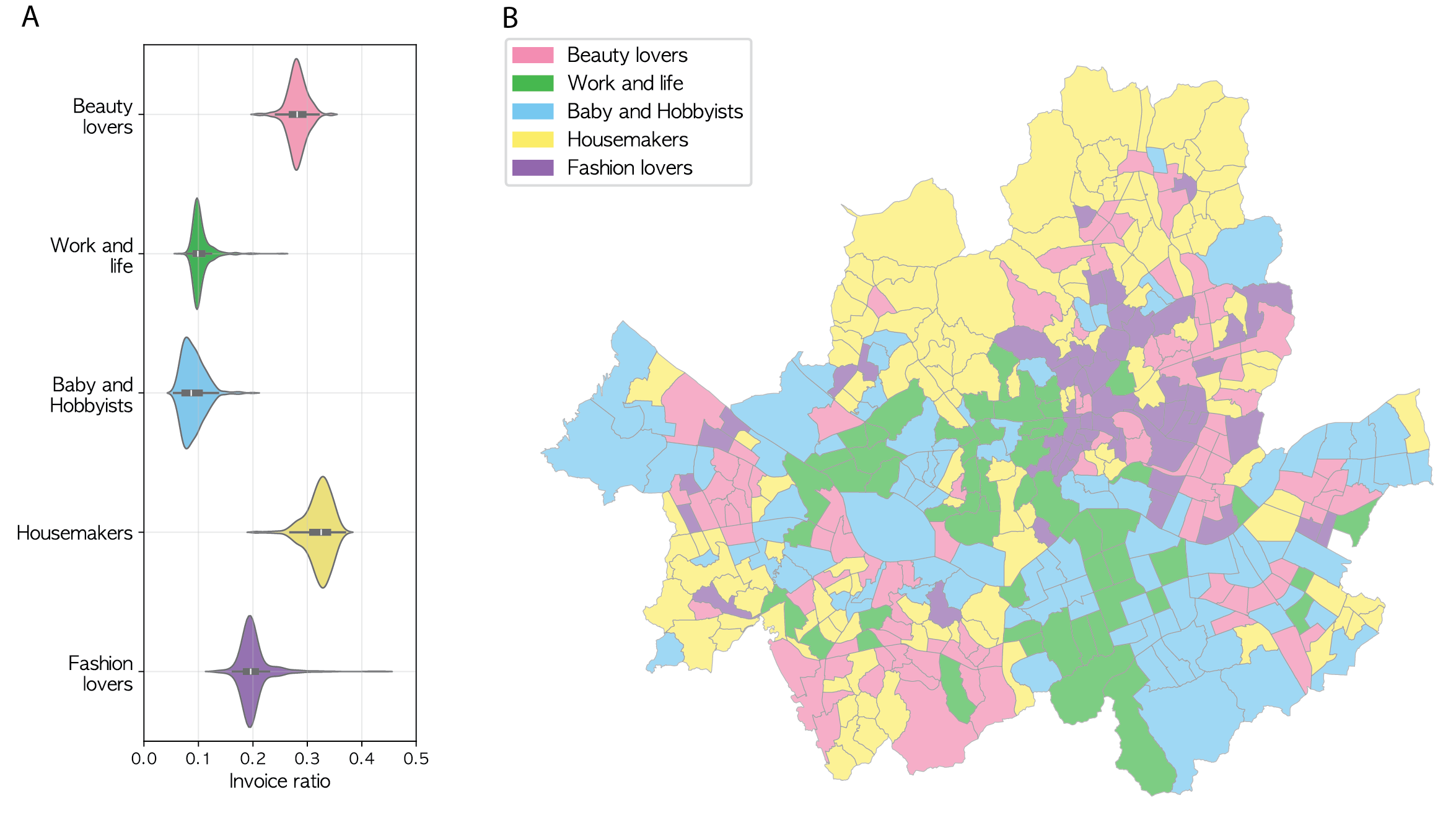}
\caption{{\bf Lifestyle composition.} A: Distribution ratio of invoice counts for products in each lifestyle cluster. B: Representative lifestyles at the district level.}
\label{fig4}
\end{figure}

The lifestyle clusters identified in the previous section can be considered as the core components that shape an urban lifestyle. To understand the overall structure of lifestyle patterns, we calculate the proportional composition of these clusters, which is the relative invoice count of each cluster within the consumption profile. We analyze these lifestyle compositions at the district level (\textit{Dong} in Korean). Thus, the lifestyle compositions are described as the invoice ratio \( r_{d,c} \) of cluster \( c \) in district \( d \) as:

\begin{equation}
r_{d,c} = \frac{I_{d,c}}{\sum_{k=1}^{C} I_{d,k}}
\end{equation}
 where \( I_{d,c} \) is the number of delivery invoices for cluster \( c \) in district \( d \), and \( C \) is the total number of lifestyle clusters.
In Fig.~\ref{fig4}A, we present the distribution of \( r_{d,c} \) across all districts for each cluster. The Poisson-like distribution suggests that lifestyle compositions are not highly diverse at the district level but follow a general pattern. On average, people tend to consume more products from the \textit{Housemakers}, \textit{Beauty lovers}, and \textit{Fashion lovers} clusters, while products related to \textit{Work and Life} and \textit{Baby and Hobbyist} are consumed less frequently. However, within this general pattern, some districts exhibit substantial variation, where certain lifestyle clusters appear in significantly higher proportions than the average. 

A particular lifestyle cluster dominating a regional consumption pattern can be used to define the region.
In Fig. \ref{fig4}B, we illustrate the representative lifestylewfor each district by standardizing the invoice ratios \( z_{d,c}\) and identifying the lifestyle cluster with the highest standardized value. The results indicate that representative lifestyles exhibit a somewhat complex spatial pattern but also reveal clusters of similar lifestyles across regions. For instance, western-northern districts predominantly exhibit \textit{Housemakers} lifestyles, whereas eastern-southern districts are characterized by \textit{Work and Life} and \textit{Baby and Hobbyists} lifestyles. This pattern suggests a potential relationship between lifestyle and the socio-economic characteristics of these areas. Northern districts are primarily residential areas with fewer business-related facilities, whereas southern districts contain major business hubs and affluent residential zones, contributing to lifestyle distinctions in these regions.

To further understand the relationship between lifestyle composition and socio-economic characteristics, we investigate how lifestyle clusters are associated with key demographic and economic indicators, such as age\cite{sgis2021gridpop}, household type\cite{kostat2023householdtype}, life cycle\cite{kostat2023lifecycle}, and income level\cite{seoul2023commercial}. Our regression analysis provides a detailed understanding of these relationships. The results in Table \ref{tab:regression} illustrate several key patterns. First, as shown in Columns (1) and (5), income level is negatively associated with the proportion of \textit{Beauty lovers} and \textit{Fashion lovers} lifestyles, suggesting that lower-income areas tend to prioritize beauty and Fashion related lifestyles more than higher-income areas. Conversely, Columns (2) and (3) suggest that higher-income districts exhibit greater spending on \textit{Work and Life} and \textit{Baby and Hobbyist} products. This indicates that districts with higherwincomes have more spending on work-related, family-oriented, and leisure-related goods. 

The coefficients of variable \textit{log(Single household)} and \textit{log(Average age)} also highlight the role of household composition. Single households are positively correlated with \textit{Beauty lovers} and \textit{Work and Life} lifestyles, while negatively correlated with \textit{Baby and Hobbyists} and \textit{Housemakers} lifestyles. This suggests that areas with a higher proportion of single-person households are more likely to focus on self-care and work-related consumption. In contrast, family-oriented lifestyles are more prevalent in areas with lower single-household rates. Additionally, older populations are strongly associated with \textit{Housemakers} lifestyles, whereas younger populations are more likely to be associated with \textit{Beauty lovers} and \textit{Baby and Hobbyists} lifestyles.

Our findings suggest that lifestyle cannot be fully explained by socio-economic characteristics alone. Even when we analyze lifestyle composition through key socio-economic indicators, the relationship remains only partial, indicating that lifestyle formation involves more than just structural factors. Instead, lifestyle emerges from a more complex interplay of various elements, such as individual preferences, cultural contexts, and behavioral tendencies, which cannot be fully captured by a combination of demographic and economic variables alone. That is, lifestyle is not just an outcome of structural conditions but also shaped by broader cultural and personal contexts that influence consumption behaviors in unique ways.

\begin{table}[!htbp] 
\centering 
\caption{Comparison of Regression Models} 
\label{} 
\resizebox{\textwidth}{!}{%
\begin{tabular}{@{\extracolsep{5pt}}lccccc} 
\\[-1.8ex]\hline 
\hline \\[-1.8ex] 
 & \multicolumn{5}{c}{\textit{Dependent variable: Invoice Ratio, \( r_{d,c}\)}} \\ 
\cline{2-6} 
\\[-1.8ex] & Beauty lovers & Work and life & Baby and Hobbyists & Housemakers & Fashion lovers \\
\\[-1.8ex] & (1) & (2) & (3) & (4) & (5)\\ 
\hline \\[-1.8ex] 
 \textit{log(Population)} & $-$0.001 & $-$0.016$^{***}$ & 0.002 & 0.017$^{***}$ & 0.003 \\ 
  & (0.002) & (0.003) & (0.002) & (0.004) & (0.006) \\ 
  & & & & & \\ 
 \textit{log(Income)} & $-$0.030$^{***}$ & 0.024$^{***}$ & 0.037$^{***}$ & $-$0.016$^{***}$ & $-$0.020$^{**}$ \\ 
  & (0.003) & (0.003) & (0.002) & (0.005) & (0.008) \\ 
  & & & & & \\ 
 \textit{log(Birth)} & $-$0.001 & 0.002 & 0.001 & 0.002 & $-$0.004 \\ 
  & (0.001) & (0.002) & (0.001) & (0.002) & (0.004) \\ 
  & & & & & \\ 
 \textit{log(Single household)} & 0.008$^{***}$ & 0.012$^{***}$ & $-$0.006$^{***}$ & $-$0.020$^{***}$ & 0.002 \\ 
  & (0.001) & (0.001) & (0.001) & (0.002) & (0.003) \\ 
  & & & & & \\ 
 \textit{log(Average age)} & $-$0.059$^{***}$ & $-$0.028$^{**}$ & $-$0.057$^{***}$ & 0.136$^{***}$ & 0.012 \\ 
  & (0.011) & (0.013) & (0.009) & (0.020) & (0.029) \\ 
  & & & & & \\ 
 \textit{Constant} & 0.872$^{***}$ & $-$0.105 & $-$0.219$^{***}$ & 0.016 & 0.407$^{*}$ \\ 
  & (0.082) & (0.094) & (0.069) & (0.146) & (0.213) \\ 
  & & & & & \\ 
\hline \\[-1.8ex] 
Observations & 374 & 374 & 374 & 374 & 374 \\ 
Adjusted R$^{2}$ & 0.488 & 0.275 & 0.741 & 0.339 & 0.053 \\ 
Residual Std. Error (df = 368) & 0.010 & 0.011 & 0.008 & 0.018 & 0.026 \\ 
F Statistic (df = 5; 368) & 71.972$^{***}$ & 29.310$^{***}$ & 214.915$^{***}$ & 39.234$^{***}$ & 5.190$^{***}$ \\ 
\hline 
\hline \\[-1.8ex] 
\textit{Note:}  & \multicolumn{5}{r}{$^{*}$p$<$0.1; $^{**}$p$<$0.05; $^{***}$p$<$0.01} \\ 
\end{tabular}%
} 
\label{tab:regression}
\end{table}

\subsection*{Temporal pattern of lifestyle between 2018 and 2022}

\begin{figure}[!h]
\centering
\includegraphics[width=13cm]{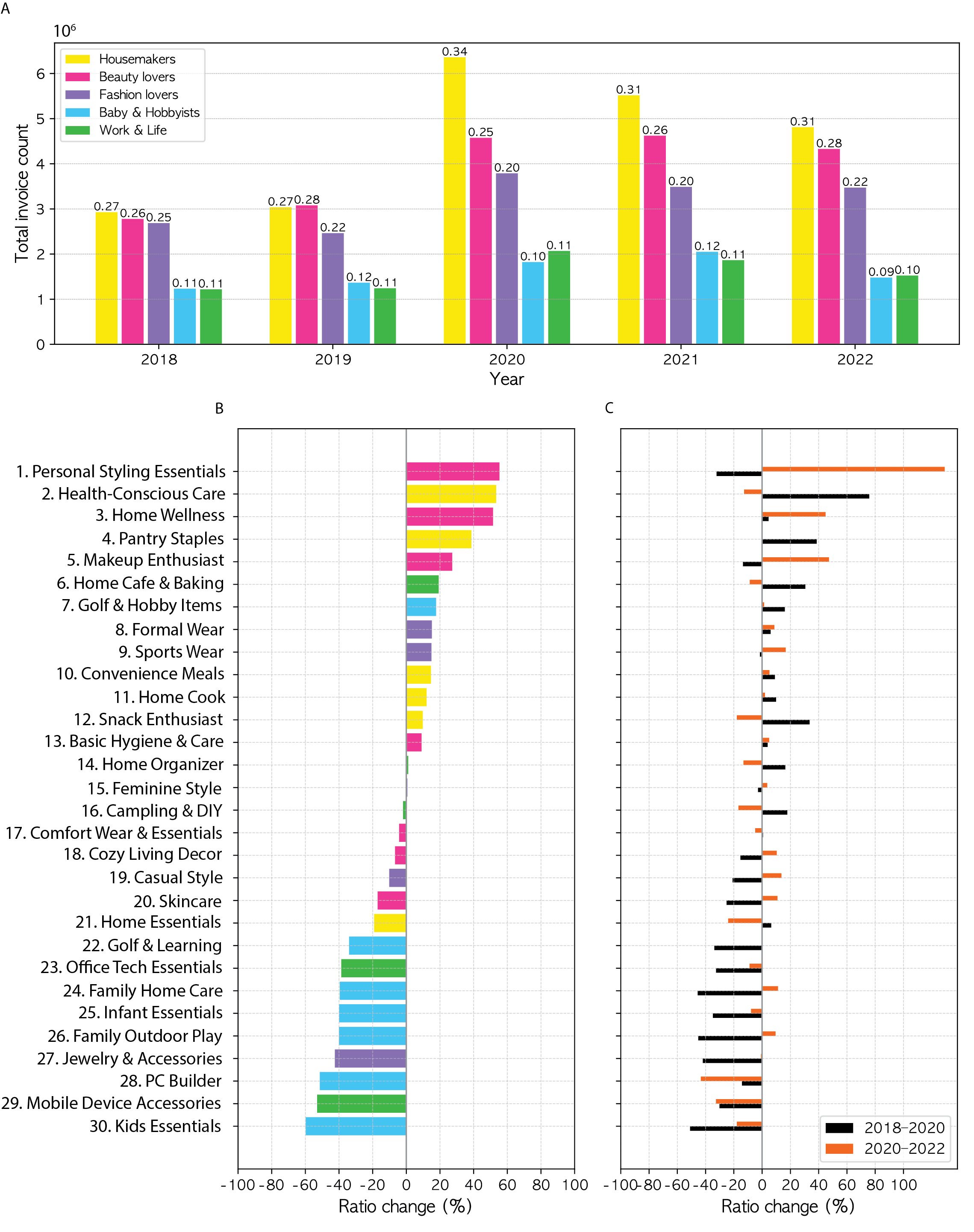}
\caption{{\bf Temporal change of lifestyles} A: Temporal trends in total invoice counts by major lifestyle clusters from 2018 to 2022. The invoice ratio for each cluster is labeled on the bars. B: Changes in invoice ratios over time for lifestyle sub-clusters. The bar colors match the large lifestyle clusters shown in panel A. C: Invoice ratio changes across two distinct periods: 2018–2020 (black lines) and 2020–2022 (orange lines), illustrating shifts in sub-cluster consumption patterns}
\label{fig5}
\end{figure}

Lastly, we analyze temporal trends in the proportion of the five lifestyles over the five-year period from 2018 to 2022. Fig \ref{fig5}A shows that the order of lifestyle composition has not changed, but the proportion itself has changed during the period. The most notable change is a sharp increase in \textit{Housemakers} lifestyles in 2020 when COVID-19 embarked. The increased concentration on \textit{Housemakers} lifestyle becomes declined after 2020, but does not return to the level before 2020, indicating the long-term effect of the pandemic on the urban lifestyle.

Using the second and third levels of communities, we identified more specific patterns in the evolution of lifestyle. As shown in Fig \ref{fig5}B, changes in lifestyle ratios widely range from a 60$\%$ decline to a 60\% increase. Note that the sub-clusters presented in Fig.~\ref{fig5}B and Fig.~\ref{fig5}C largely follow the classification structure established in Fig.~\ref{fig3}. However, sub-clusters containing fewer than five product items were either excluded from the detailed temporal analysis or aggregated to their higher-level community to ensure statistical robustness and interpretability.

The overall shifts observed over these five years can be categorized into three major trends: 1) The rising focus on personal well-being and home-centered activities, 2) The decline of IT and technology-related lifestyles, 3) The diminishing prevalence of family- and child-related lifestyles. 
First, we observed a clear and consistent rise in self-oriented lifestyle clusters, particularly those centered on personal care and well-being. The most significantly increasing sub-clusters include personal styling essentials, health-conscious care, home wellness, pantry staples, and makeup-related items (1, 2, 3, 4, 5, in Fig.~\ref{fig3}B and C). These trends point to a growing emphasis on self-care, appearance, and overall wellness—shaped and reinforced during the pandemic—that has persisted beyond the initial crisis. While home-centered activities such as cooking and interior improvement (3, 10, 11) also increased, their growth was less dramatic. This indicates that the dominant shift was not solely about staying indoors, but about investing in the self within the home.

On the other hand, among the nine sub-clusters that experienced more than a 30 percent decrease, four (22, 23, 28, 29) were associated with computers and mobile IT devices. Despite the increased reliance on remote work and online education during the pandemic, expenditures on computers, office supplies, and digital hardware sharply declined. This trend, which began even before the pandemic, suggests a broader shift in digital lifestyles—possibly moving toward a mobile- and cloud-based culture rather than traditional PC usage. A similarly consistent decline was observed in family- and child-related consumption patterns (24, 25, 26, 30), which appears to reflect Korea’s ongoing low birth rate and the shift away from parenting as a normative life course. 

While the overall trends observed over the five-year period point to structural shifts, some lifestyle sub-clusters displayed short-term fluctuations in response to the COVID-19 crisis. For instance, activities that could be enjoyed with minimal risk of infection—such as camping, gardening, and snack consumption (12, 14, 16)—experienced considerable growth between 2018 and 2020, during the early phase of the pandemic. However, as pandemic-related fears gradually subsided and mobility restrictions were lifted by 2022, demand for these items relatively declined. Conversely, outdoor-related categories such as sportswear and family outdoor play (9, 26), which had declined during the pandemic, rebounded as people resumed more active lifestyles. These fluctuation patterns suggest that some behavioral changes were temporary reactions to external constraints, while others may have evolved into more permanent lifestyle elements.

\section*{Discussion}
Lifestyle serves as a framework for understanding a society and its members, encompassing their social status, consumption patterns, values, and cultural preferences. With the growing availability of large-scale purchasing data, such as credit card transactions, data-driven research has increasingly explored lifestyles through consumer behavior. However, the absence of detailed information on individual purchases poses a challenge in accurately capturing the structure of lifestyles based on consumption patterns.

In this study, we proposed a novel approach to analyzing lifestyle structures by leveraging detailed product-level consumption data. By constructing a co-consumption network, we systematically identified meaningful product groupings, revealing the underlying components that shape urban lifestyles. This data-driven classification offers a hierarchical framework that captures both five broad lifestyle categories and fine-grained insights into specific consumption patterns. Our findings suggest that lifestyle is a structured and dynamic system, emerging from the interactions between consumer preferences, social environments, and broader economic and cultural influences. 

The hierarchical analysis demonstrated that urban lifestyles consist of distinct yet interconnected clusters, with \textit{Housemakers}, \textit{Beauty lovers}, and \textit{Fashion lovers} being the most dominant. Meanwhile, \textit{Baby and Hobbyists} exhibited the most distinctive patterns, both in network structure and spatial distribution. This reflects a recent trend in Korea, where childbirth and child-rearing are no longer part of a universal life course, but rather resemble selective lifestyle choices—similar to how people choose to engage in hobbies.

Furthermore, we examined the relationship between lifestyle compositions and socioeconomic factors, highlighting that while income, household structure, and age demographics partially explain lifestyle distributions, they do not fully capture the complexity of lifestyle formation. Our temporal analysis further demonstrated that lifestyle patterns are not static but instead evolve in response to external disruptions. The COVID-19 pandemic and social structural changes significantly altered consumption behaviors, leading to a surge in home-centered lifestyles, increased demand for self-care and wellness products, and a decline in technology and parenting-related expenditures. While some of these shifts were temporary, others—such as the growing emphasis on home improvement and personal well-being—appear to have solidified as long-term lifestyle changes.

Unlike traditional consumption analysis methods that rely on predefined categories or aggregate statistics, our network-based approach uncovers emergent lifestyle structures derived from actual co-consumption behavior. This enables a more data-driven and contextually grounded understanding of urban lifestyles. The hierarchical and spatially structured nature of lifestyle clusters --- revealed through network topology --- provides insights that would not be accessible through linear or category-based analyses alone. While co-consumption analysis is also widely used in the retail industry, such as in recommendation systems by retail companies --- these applications primarily focus on individual-level prediction and personalization for commercial purposes. In contrast, our study leverages co-consumption networks to reveal the structural organization of lifestyle communities, their spatial characteristics, and their collective socioeconomic meanings.

Our study has several limitations that must be acknowledged.
First, although online consumption is now widespread and accounts for a substantial portion of total consumer spending in Seoul, it does not encompass all forms of consumption. In particular, due to disparities in accessibility and availability of online shopping platforms, our data may exhibit inherent biases particularly across age groups and product categories. Second, our analysis specifically focuses on product-based consumption patterns as a way to define and classify lifestyle. As discussed in the introduction, lifestyle is a broad and multifaceted concept that encompasses various dimensions of an individual's life --- including values, social identity, cultural norms, and behavioral routines. Therefore, we do not claim that our analysis captures lifestyle in its entirety. Rather, it should be understood as an attempt to uncover one important dimension of lifestyle through the lens of consumption behavior. To address these limitations, future research could incorporate additional data sources to construct a more comprehensive picture of lifestyle. For example, combining our package delivery dataset with offline consumption records---such as point-of-sale (POS) data or receipt-level purchase information---would enable a more complete understanding of overall consumption behavior. Moreover, integrating mobility data or datasets that reflect individuals' cultural tastes, values, and media consumption habits could further enrich the analysis, allowing researchers to capture lifestyle patterns beyond material consumption.

There are also promising directions for extending the framework beyond the current scope. Our results suggest that lifestyle may be understood not only as a socially constructed concept, but also as a phenomenon embedded within spatial contexts. As seen in Fig. \ref{fig4}B, lifestyles appear to be spatially clustered, raising questions about the mechanisms of spatial homophily in consumption behavior. Prior studies on urban lifestyle formation suggest that lifestyles evolve under the influence of neighbors, reinforcing shared consumption behaviors and shaping new cultural patterns~\cite{boy2020instagram, bellah1985habits, florida2002rise}. The concept of ``lifestyle enclaves" describes the emergence of social clusters where individuals share similar lifestyles through aesthetic preferences, leisure activities, and consumption habits~\cite{zukin1995cultures, low2003gated}. In contemporary urban environments, these enclaves are no longer strictly residential communities but are also fluid and mobile, emerging through consumption practices in shared urban spaces such as high-end cafes, fitness clubs, and specialized retail hubs. Thus, future studies could apply spatial clustering techniques to more deeply analyze the spatial dimension of lifestyle formation. This would allow us to investigate how lifestyle enclaves are formed in urban areas, how they change over time, and whether certain lifestyle groups exhibit stronger spatial homophily than others. Moreover, longitudinal extensions of this work could explore lifestyle trajectories, revealing how individuals or neighborhoods shift between lifestyle categories in response to social, economic, or environmental changes.

Our study contributes to a more comprehensive understanding of lifestyle structures, offering valuable insights for urban policy, market segmentation, and consumer behavior analysis. By capturing consumption behaviors at a highly granular level, our bottom-up approach provides a scalable and replicable framework for analyzing lifestyle patterns in different urban contexts. As urban societies continue to evolve, our framework can serve as a powerful tool for researchers, policymakers, and businesses seeking to understand the shifting dynamics of contemporary lifestyles.

\section*{Materials and methods}
\subsection*{Data description and data processing}

Our dataset comprises 103,342,186 package delivery records provided by CJ Olive Networks, a major Korean logistics company. It includes all delivery records sent to Seoul during the month of June from 2018 to 2022, with yearly counts as follows:  
16,835,330 (2018), 16,577,235 (2019), 25,004,774 (2020), 23,223,759 (2021), and 21,701,088 (2022). 
The dataset was constructed by digitizing package invoices through OCR (optical character recognition). Product categories are structured in a three-level hierarchy, comprising 10 large-level categories, 128 mid-level categories, and 534 small-level categories (Detailed category names are provided in S1 Table).
To ensure privacy, individual daily records are aggregated at the monthly level and spatially grouped into 50 × 50-meter grid cells. Each spatial unit typically covers one or two apartment buildings and a few low-rise houses within Seoul.

Since the dataset contains detailed information at a highly granular level, we apply filtering procedures to reduce noise and focus on personal consumption behavior. First, to eliminate commercial-purpose deliveries, we identify and exclude non-residential areas where the registered residential population is below a threshold of 20 persons\cite{kostat2025popguide}. Additionally, we remove grid cells that received fewer than 500 delivery invoices, as well as product categories with fewer than 10,000 deliveries across the entire dataset. After this filtering process, the number of valid grid cells is reduced from 103,804 to 92,122, and the number of small-level product categories is reduced from 534 to 453.

Our analysis using package delivery data is particularly relevant in the context of Korean cities, where online shopping accounts for 27\% of total consumer purchases \cite{kostat2025online,statista2023categories}. Moreover, nearly all types of goods, including fresh food, are available for purchasing online, making e-commerce a dominant and comprehensive retail channel. Unlike in some regions where online shopping is concentrated in specific product categories, in Korea, the share of online shopping is evenly distributed across all product types\cite{kostat2020social}. This widespread adoption ensures that package delivery data reflects a well-balanced and representative picture of consumer behavior across various product domains, making it a powerful resource for analyzing lifestyle patterns.

\subsection*{Network Filtering}

The purpose of the filtering step is to uncover the underlying community structure embedded within the network by removing noisy or insignificant edges while preserving statistically meaningful connections. To achieve this, we apply the disparity filtering technique proposed by Serrano et al.~\cite{sarrano2009filtering}, which prunes the network while maintaining its multiscale backbone. This method assesses the significance of each edge weight relative to the local connectivity structure of a node.

In disparity filtering, for a node \( i \) with \( k_i \) connections, the weight \( w_{ij} \) of an edge connecting node \( i \) to node \( j \) is normalized by the total strength \( s_i \) of node \( i \), defined as:

\begin{equation}
p_{ij} = \frac{w_{ij}}{s_i}, \quad s_i = \sum_{j} w_{ij}
\end{equation}

Under the assumption that edge weights are randomly distributed, the expected probability distribution of \( p_{ij} \) follows:

\begin{equation}
P(p_{ij}) = (k_i - 1) (1 - p_{ij})^{k_i - 2}
\end{equation}

This formulation allows for a statistical significance test for each edge. An edge is retained in the filtered network if:

\begin{equation}
1 - (k_i - 1) \int_{0}^{p_{ij}} (1 - x)^{k_i - 2} dx < \alpha
\end{equation}

where \( \alpha \) is the significance threshold. Edges that do not satisfy this condition are removed. By adjusting \( \alpha \), we control the stringency of the filtering process, balancing between eliminating noise and preserving meaningful connections.

To determine an appropriate filtering threshold, we examine the modularity of the resulting network structure across various values of \( \alpha \). We select \( \alpha = 0.29 \), which achieves an optimal balance between node retention and the interpretability of the resulting community structure. In our case, we set a modularity threshold of 0.3 to ensure that the extracted lifestyle communities are coherent~\cite{clauset2004finding, chodrow2020hypergraph}. After applying the filtering, the network is reduced from a fully connected graph of 453 nodes to a sparser network with 322 nodes and 4,824 edges.

\vspace{1em}

\subsection*{Optimized Hierarchical Community Structure of the Network}

\begin{algorithm}[!t]
\caption{Hierarchical Community Detection}
\label{alg:hierarchical}
\begin{algorithmic}[1]
\Require Graph $G = (V, E)$, modularity threshold $Q_{\text{threshold}} = 0.25$, iterations $T$
\Ensure Hierarchical community structure of $G$

\Function{StabilizedLouvain}{$G$, $T$}
    \For{$t = 1$ to $T$}
        \State Run Louvain Algorithm on $G$ and store community assignment $C^t_i$
    \EndFor
    \State For each node $i$, compute frequency $f_{i,c}$ over $T$ runs
    \State Assign node $i$ to the community with highest $f_{i,c}$
    \State \Return final community assignment
\EndFunction

\Function{FilterWeakEdges}{$G$}
    \State Remove weakly connected edges based on disparity filtering
    \State \Return filtered graph
\EndFunction

\State // Step 1: Initial Community Detection
\State $C_0 \gets$ \Call{StabilizedLouvain}{$G$, $T$}

\State // Step 2: Community Detection on Sub-Networks
\For{each community $C$ in $C_0$}
    \State Extract sub-network $G_C = (V_C, E_C)$ from $G$
    \State Compute modularity $Q_C$

    \While{$Q_C < Q_{\text{threshold}}$}
        \State $G_C \gets$ \Call{FilterWeakEdges}{$G_C$}
        \State Compute updated modularity $Q_C$
    \EndWhile

    \State $C_C \gets$ \Call{StabilizedLouvain}{$G_C$, $T$}
    \State Store $C_C$ as refined community structure for $G_C$
\EndFor

\State \Return Hierarchical community structure
\end{algorithmic}
\end{algorithm}

We detect community structures in the co-consumption network by optimizing modularity, which measures the strength of division of a network into communities. The modularity \( Q \) of a partition is given by:

\begin{equation}
Q = \frac{1}{2m} \sum_{i,j} \left[ A_{ij} - \frac{k_i k_j}{2m} \right] \delta(c_i, c_j),
\end{equation}
where \( A_{ij} \) represents the weight of the edge between nodes \( i \) and \( j \), \( k_i \) and \( k_j \) denote the sum of edge weights for nodes \( i \) and \( j \), \( m \) is the total edge weight in the network, and \( \delta(c_i, c_j) \) is the Kronecker delta function that equals 1 if nodes \( i \) and \( j \) are in the same community and 0 otherwise.

To optimize modularity, we employ the Louvain algorithm, which efficiently identifies densely connected groups by iteratively maximizing \( Q \). Given the stochastic nature of the algorithm, we repeat the detection process \( T \) times to ensure stability~\cite{lee2021inconsistence, Cho2023}. Each node is then assigned to the most frequently detected community across iterations. For each node \( i \), we define the probability of belonging to community \( c \) as:

\begin{equation}
f_{i,c} = \frac{1}{T} \sum_{t=1}^{T} \delta(C_i^t, c),
\end{equation}

 where \( C_i^t \) represents the community assigned to node \( i \) at iteration \( t \). By computing \( f_{i,c} \), we estimate the most stable community assignment for each node and mitigate the effects of randomness in the Louvain algorithm.

Furthermore, we explore hierarchical structure by detecting sub-communities within the initially identified communities. We apply the same filtering-detection process recursively to these sub-networks~\cite{park2019global}. Specifically, for each community, we extract its corresponding sub-network and recalculate modularity. If the modularity remains below the threshold, we iteratively apply the filtering process until the modularity exceeds the predefined level. This hierarchical and iterative detection process enables us to construct a reliable multi-level representation of lifestyle clusters in the consumption network.

\section*{Supporting information}






\paragraph*{S1 Table.}
\label{S1_Table}
{\bf Full list of large and small categories.} It provides the complete list of large and small category classifications used in the raw dataset provided by CJ Olive Networks.

\section*{Acknowledgments}
This work was supported by the National Research Foundation of Korea(NRF) grant funded by the Korea government(MSIT)(RS-2022-NR070854) and KDI School of Public Policy and Management (20240049).



\begin{thebibliography}{10}

\bibitem{zukin1998urban}
Zukin S.  
\newblock Urban lifestyles: Diversity and standardisation in spaces of consumption.  
\newblock Urban Studies. 1998;35(5–6):825–839.

\bibitem{bourdieu1984distinction}
Bourdieu P.  
\newblock Distinction: A Social Critique of the Judgement of Taste.  
\newblock Harvard University Press; 1984.

\bibitem{bellah1985habits}
Bellah RN, Madsen R, Sullivan WM, Swidler A, Tipton SM.  
\newblock Habits of the Heart: Individualism and Commitment in American Life.  
\newblock University of California Press; 1985.

\bibitem{kozlowski2019}
Kozlowski AC, Taddy M, Evans JA.  
\newblock The geometry of culture: Analyzing the meanings of class through word embeddings.  
\newblock Am Sociol Rev. 2019;84(5):905–949.

\bibitem{Posta2015latte}
DellaPosta D, Shi Y, Macy MW.  
\newblock Why do liberals drink lattes?  
\newblock American Journal of Sociology. 2015;120(5):1473–1511.

\bibitem{shi2017}
Shi F, Shi Y, Dokshin FA, Evans JA, Macy MW.  
\newblock Millions of online book co-purchases reveal partisan differences in the consumption of science.  
\newblock Nat Hum Behav. 2017;1:0079.

\bibitem{Praet2021lifestylepoliticsmodeling}
Praet S, Van Aelst P, van Erkel P, Van der Veeken S, Martens D.  
\newblock Predictive modeling to study lifestyle politics with Facebook likes.  
\newblock EPJ Data Science. 2021;10(1):50.

\bibitem{boy2020instagram}
Boy JD, Uitermark J.  
\newblock Lifestyle enclaves in the Instagram city?  
\newblock Social Media + Society. 2020;6(3):1-11.

\bibitem{low2003gated}
Low S.  
\newblock Behind the Gates: Life, Security, and the Pursuit of Happiness in Fortress America.  
\newblock Routledge; 2003.

\bibitem{florida2002rise}
Florida R.  
\newblock The Rise of the Creative Class.  
\newblock Basic Books; 2002.

\bibitem{zukin1995cultures}
Zukin S.  
\newblock The Cultures of Cities.  
\newblock Blackwell; 1995.

\bibitem{di2018sequences}
Di Clemente R, Luengo-Oroz M, Travizano M, Xu S, Vaitla B, González MC.  
\newblock Sequences of purchases in credit card data reveal lifestyles in urban populations.  
\newblock Nat Commun. 2018;9(1):3330.

\bibitem{hidalgo2007product}
Hidalgo CA, Klinger B, Barabási AL, Hausmann R.  
\newblock The product space conditions the development of nations.  
\newblock Science. 2007;317(5837):482–487.

\bibitem{neffke2019value}
Neffke FMH.  
\newblock The Value of Complementary Co-workers.  
\newblock Science Advances. 2019;5(12):eaax3370.  
\newblock doi:10.1126/sciadv.aax3370.


\bibitem{newman2006modularity}
Newman MEJ.  
\newblock Modularity and community structure in networks.  
\newblock Proceedings of the National Academy of Sciences. 2006;103(23):8577--8582.

\bibitem{blondel2008fast}
Blondel VD, Guillaume J-L, Lambiotte R, Lefebvre E.  
\newblock Fast unfolding of communities in large networks.  \newblock J Stat Mech. 2008;2008(10):P10008.

\bibitem{lee2021inconsistence}
Lee D, Lee SH, Kim BJ, Kim H.  
\newblock Consistency landscape of network communities.  
\newblock Physical Review E, 2021;103(5):052306.

\bibitem{Cho2023}
Cho W, Lee D, Kim BJ.  
\newblock A multiresolution framework for the analysis of community structure in international trade networks.  
\newblock Scientific Reports. 2023;13(1):5721.

\bibitem{park2019global}
Park J, Wood IB, Jing E, Nematzadeh A, Ghosh S, Conover MD, Ahn YY.  
\newblock Global labor flow network reveals the hierarchical organization and dynamics of geo-industrial clusters.  
\newblock Nature Communications. 2019;10(1):3449.

\bibitem{sgis2021gridpop}
Statistics Korea  
\newblock Grid Statistics (Population).
\newblock Statistical Geographic Information Service; 2021. Available from: \url{https://sgis.kostat.go.kr/view/pss/openDataIntrcn}

\bibitem{kostat2023householdtype}
Statistics Korea.  
\newblock Statistics of Households in Seoul by Household type in 2023.  
\newblock Daejeon: Statistics Korea; 2023
\newblock Available from: \url{https://kosis.kr/statHtml/statHtml.do?orgId=101&tblId=DT_1JC1501&conn_path=I2}

\bibitem{kostat2023lifecycle}
Statistics Korea.  
\newblock Number of Vital Events (Births, Deaths, Marriages, Divorces) in 2023.  
\newblock Daejeon: Statistics Korea; 2023.  
\newblock Available from: \url{https://kosis.kr/statHtml/statHtml.do?orgId=101&tblId=DT_1B8000K&conn_path=I2}

\bibitem{seoul2023commercial}
Seoul Metropolitan Government.  
\newblock Seoul Commercial District Analysis Service (Income and Consumption by Administrative District).  
\newblock Seoul Open Data Plaza; 2023. Available from: \url{https://data.seoul.go.kr/dataList/OA-22166/S/1/datasetView.do}

\bibitem{kostat2025popguide}
Statistics Korea.  
\newblock Geographic Information System statistic guide.
\newblock Daejeon: Statistics Korea; 2025.  
\newblock Available from: \url{https://sgis.kostat.go.kr/contents/include/download.jsp?filename=GIS_statistics_guid.zip&path=/board/&type=board}


\bibitem{kostat2025online}
Statistics Korea.  
\newblock Online Shopping Trends in January 2025 (Press Release).  
\newblock Daejeon: Statistics Korea; 2025.  
\newblock Available from: \url{https://kostat.go.kr/board.es?mid=a10301120300&bid=241&act=view&list_no=435317}


\bibitem{statista2023categories}
Statista.  
\newblock Most popular product categories in social commerce in South Korea as of August 2023.  
\newblock Statista Research Department; 2023.  
\newblock Available from: \url{https://www.statista.com/statistics/314776/most-popular-social-commerce-product-categories-in-south-korea/}

\bibitem{kostat2020social}
Statistics Korea.  
\newblock Korea Social Trends 2020.  
\newblock Daejeon: Statistics Korea; 2020.  
\newblock Available from: \url{https://kostat.go.kr/boardDownload.es?bid=12308&list_no=386918&seq=2}

\bibitem{sarrano2009filtering}
Serrano MA, Boguná M, Vespignani A.  
\newblock Extracting the multiscale backbone of complex weighted networks.  
\newblock Proc Natl Acad Sci USA. 2009;106(16):6483–6488.

\bibitem{clauset2004finding}
Clauset A, Newman MEJ, Moore C.  
\newblock Finding community structure in very large networks.  
\newblock Phys Rev E. 2004;70(6):066111.

\bibitem{chodrow2020hypergraph}
Chodrow PS.  
\newblock Hypergraph clustering by iteratively reweighted modularity maximization.  
\newblock Sci Adv. 2020;6(19):eaay6373.



\end{thebibliography}
\end{document}